\documentstyle[psfig]{aa}
\def\gsimeq
{\hbox{\raise0.5ex\hbox{$>\lower1.06ex\hbox{$\kern-1.07em{\sim}$}$}}}
\def\lsimeq
{\hbox{\raise0.5ex\hbox{$<\lower1.06ex\hbox{$\kern-1.07em{\sim}$}$}}}

\begin{document}
   \title{X-ray evidence for a mildly relativistic and variable outflow in the luminous Seyfert 1 galaxy Mrk~509}


   \author{M. Cappi
\inst{1}
\and
F. Tombesi
\inst{1,2,3,4}
\and
S. Bianchi
\inst{5}
\and
M. Dadina
\inst{1}
\and
M. Giustini
\inst{1,2,6}
\and
G. Malaguti
\inst{1}
\and
L. Maraschi
\inst{7}
\\
G.G.C. Palumbo
\inst{2}
\and
P.O. Petrucci
\inst{8}
\and
G. Ponti
\inst{9}
\and
C. Vignali
\inst{2}
\and
T. Yaqoob
\inst{3,4}
}

   \offprints{M. Cappi \\ \email{cappi@iasfbo.inaf.it}}

   \institute{INAF-IASF Bologna, Via Gobetti 101, I-40129 Bologna, Italy 
\and
Dipartimento di Astronomia, Universit\`a degli Studi di Bologna, Via Ranzani 1, I-40127 Bologna, Italy 
\and
Department of Physics and Astronomy, Johns Hopkins University, 3400, Baltimore, MD 21218, USA
\and
Laboratory for High Energy Astrophysics, NASA/Goddard Space Flight Center, Greenbelt, MD 20771, USA
\and
Dipartimento di Fisica, Universit\`a degli Studi Roma Tre, Via della Vasca Navale 84, 00146, Roma, Italy
\and
Department of Astronomy \& Astrophysics, Pennsylvania State University, University Park, PA 16802, USA
\and 
INAF-Osservatorio Astronomica di Brera, Via Brera 28, 20121, Milano, Italy
\and
Laboratoire d'Astrophysique de Grenoble, Universit\'e Joseph-Fourier/CNRS UMR 5571, 38041, Grenoble, France
\and 
APC Universit\'e Paris 7 Denis Diderot, 75205, Paris, France
}

\date{Received / Accepted}

\markboth{Cappi et al.: High velocity and variable iron absorption in Mrk~509}{}
\titlerunning{High velocity and variable iron absorption in Mrk~509}


 \abstract
{There is growing evidence for the presence of blueshifted Fe K absorption lines 
in a number of radio-quiet AGNs and QSOs. These may be fundamental to probe flow dynamics 
near supermassive black holes.}
{Here we aim at verifying and better characterising 
the existence of such Fe K absorption at $\sim$8-10~keV in the luminous Seyfert 1 galaxy Mrk~509, one of the 
most promising target for these studies.}
{We present a comprehensive spectral analysis of the six  \emph{XMM-Newton} observations of the source (for a total of $\sim$200 ks), 
focusing on the detailed and systematic search for absorption features in the high-energy data.}
{We detect several absorption features at rest-frame energies $\sim$8-8.5~keV and $\sim$9.7~keV.
The lines are consistent with being produced by H-like iron K$\alpha$ and K$\beta$ shell absorptions associated with 
an outflow with mildly relativistic velocity of $\sim$0.14--0.2~c. 
The lines are found to be variable in energy and, marginally in intensity, implying that variations 
in either the column density, geometry and/or ionization structure of the outflow are common in this source.}
{}

   \keywords{Galaxies: active -- X-rays: galaxies -- Galaxies: individual: \object{Mrk~509}}

   \maketitle
%

\section{Introduction}

There is compelling evidence for the presence of blueshifted Fe absorption lines 
at rest-frame energies of $\sim$ 7-10 keV in the X-ray spectra of several 
radio-quiet (RQ) AGNs and QSOs. These are commonly interpreted as 
due to resonant absorption from FeXXV and/or FeXXVI ions associated to a
zone of circumnuclear gas with log $\xi$ $\sim$ 2-4 erg~s$^{-1}$~cm and column density 
$\sim$ 10$^{22-24}$~cm$^{-2}$ (see Cappi 2006 and references therein), i.e. 
significantly higher than the absorption layers commonly associated to warm absorbers 
(Crenshaw et al. 2003, Blustin et al. 2005). 


The velocities of these absorbers being often quite large, reaching sometimes 
mildly relativistic speeds (up to 0.2-0.4 c), suggest we might be probing the 
extreme component of a wind/outflow that may well carry large amount of mass 
into the interstellar medium (ISM) and/or intergalactic medium (IGM) at a rate comparable to, or even larger than the Eddington 
accretion rate (King and Pounds 2003, Reeves et al. 2003, Chartas et al. 2003, Pounds \& Page 2006).

Most interestingly, variability of the absorption on timescales down to a few ks has been found 
among some sources (Risaliti et al. 2005, Dasgupta et al. 2005, Braito et al. 2007, Turner et al. 2007, 
Reeves et al. 2008, Saez et al. 2009, Giustini et al., in prep.).  
Variability studies could, in principle, allow us to estimate the location and geometry of the absorber 
(Nicastro et al. 1999, Krongold et al. 2007, Risaliti et al. 2005, Reeves et al.  2008), thereby placing more stringent 
limits to the remaining large (order of magnitude) uncertainties in the 
total mass, energy and momentum that is driven by these outflows (Elvis 2006).

Overall, these studies are fundamental to probe and characterise the most extreme 
(in terms of velocity, ionization state, mass outflow rate)
phases of outflows, so helping in quantifying the potential feedback impact of AGNs 
onto their host galaxies and the IGM.


Mrk~509 (z=0.034) is one of the X-ray brightest (2--10~keV 
flux of $\sim$2--5$\times10^{-11}$~erg~cm$^{-2}$~s$^{-1}$) and most luminous
(L$_{2-10{\rm keV}}$ $\sim 1-3 \times 10^{44}$ erg/s) Seyfert 1 galaxy known.
It is an interesting science case for these studies because a previous 
study by Dadina et al. (2005), based on \emph{BeppoSAX} multiple observations and on
two early \emph{XMM-Newton} observations, found a transient and blueshifted 
absorption line at a rest-frame energy of $\sim$ 8.2 keV. In addition, a transient, redshifted, 
absorption line at a rest-frame energy of $\sim$5.4 keV was found.
Motivated by these results, additional monitoring of the source 
with \emph{XMM-Newton} was then proposed and eventually performed.
Here we present the results on the systematic search and characterization of Fe {\it absorption} lines 
at high energies obtained from a comprehensive analysis on the full set of 
six \emph{XMM-Newton} observations\footnote{Results on the Fe {\it emission} line properties, for the 
same dataset, were presented in Ponti et al. 2009.}.

\section{Data Analysis and Results}

Mrk~509 was observed by the \emph{XMM-Newton} EPIC instruments on six separate 
occasions, for a total good exposure time of $\sim$280~ks (see Table 1). 
The observations were performed once in October 2000, once in 
April 2001, three times in October 2005, and once in April 2006. 
The only contiguous observations were those performed in 2005. 
The data were reduced using the standard software SAS v.~8 (de la Calle \& Loiseau N., 2008)
and the analysis was carried out using the HEASoft v.~6.5 package\footnote{http://heasarc.gsfc.nasa.gov/docs/software/lheasoft/}.
After filtering for times of high background rate, the useful exposure times, 
corrected for the live time fraction (up to 0.7 for the pn in small window mode), were between 
2.7 and 60 ks per observation (see Table \ref{tab:1}). 
The EPIC pn and MOS camera were all operated in the ``Small Window'' mode with the thin filter 
applied for all observations. 
We checked that pile-up was not significant in the pn data at this level of source flux,
even when considering both single and double events. The pile-up fraction for the MOS 
data was $\lsimeq$15\% for all observations. Also in this case, this will not affect the 
results on the narrow absorption lines presented in this work.
Source counts were extracted from a circular region of radius 40 arcsec, while 
the background counts were extracted from a nearby source-free area of 
the detector of the same size.
Given the significantly lower effective area of the MOS detector between 6-10 keV, and 
for sake of clarity and simplicity, we only report here the results obtained from the pn
data and show the MOS data for consistency checks only. 
We decide also not to report here on the observation n.~3 
because its very short exposure ($\sim$2.7 ks; see Table 1) hampered any significant 
and interesting result.

\begin{table}
\caption{Mrk 509 XMM-Newton Observations Log}
\label{tab:1}     
\begin{center}
\begin{tabular}{c c c c c}
\hline\hline             
 N$^{\circ}$ & Obs. ID & Date & Expo & Counts \\
 &   &    &   &  (2-10 keV)\\
  &   &    &  (ks) &  ($\times10^4$)\\
\hline                  
 1 & 0130720101 & 2000 Oct 25--25 & 20.7 & 7\\

 2 & 0130720201 & 2001 Apr 20--20 & 23.1 &  10\\

 3 & 0306090101 & 2005 Oct 16--16 & 2.7 & 1.1\\

 4 & 0306090201 & 2005 Oct 18--19 & 59.8 & 25\\

 5 & 0306090301 & 2005 Oct 20--20 & 32.4 & 14\\

 6 & 0306090401 & 2006 Apr 25--26 & 48.6 & 23\\
\hline                        
\end{tabular}
\end{center}
\scriptsize 
Number, Observation ID, Observation date, net EPIC pn exposure time, and 
total 2--10 keV counts for the six \emph{XMM-Newton} observations of \object{Mrk~509}.
\end{table}

\subsection{Light-curves}

Analysis of the soft and hard X-ray light curves from the six observations indicates 
only moderate ($\sim$20-30\%) flux and spectral variability on both short 
($<$ few tens ks) and long ($\sim$years) time-scales (see lightcurve in Figure 1 of Ponti et al. 2009). 
A more detailed time-resolved spectral analysis focused on the low-energy warm absorber as 
measured by the RGS is deferred to a specific paper (Detmers et al., in preparation).

\subsection{X-ray Continuum and Fe Emission Lines}

Given the weakly varying continuum, we extracted the mean pn spectra for each of the six observations,
grouping the data to a minimum of 25 counts per channel to apply the $\chi^2$ minimization 
statistics. Only the data in the 3.5--10.5~keV interval were considered, and the signal-to-noise 
ratios, even at 10 keV, were always greater than 12 (pn) and 6 (MOS).
The spectra were first fitted with a single power-law model plus a column density with value 
fixed at the Galactic one ($N_{\rm Hgal}\simeq 4\times10^{20}$~cm$^{-2}$, Dickey \& Lockman 1990).
We checked that the addition of a further neutral absorption component intrinsic to the source was not 
required by the data. 
We obtained good fits of the continuum with a typical power-law photon index of $\Gamma\simeq1.55-1.72$, but 
residuals of the data (Fig. 1, top panel) clearly showed  the presence of a neutral 
Fe~K$\alpha$ emission line at the rest frame energy of $\sim$ 6.4~keV. 
We therefore added a Gaussian emission line to the model. The line was slightly resolved 
(see also Ponti et al. 2009), and we fixed the width $\sigma$ to its best-fit value of 
$\sim$100~eV. Best-fit values with this simple model are reported in Table 2, with errors quoted at the 90\% confidence level. 
The fit improvement was of $\Delta\chi^2\geq42$ and the equivalent 
width of the emission line of the order of $\sim$60--80~eV for all the observations. 

\begin{table}[!htbp]
\caption{3.5--10.5 keV continuum and Fe emission line parameters}
\label{tab: mrk509_baseline_par}
\begin{center}
\begin{scriptsize}
\begin{tabular}{ccccccc}
\hline
N$^{\circ}$ & $\Gamma$    & Energy$^1$ & EW$^2$     & Flux$^3$   & $\chi^2/\nu$ & $\Delta\chi^2$ \\
    &             & (keV)      & (eV)& &              &                \\
\hline
1 &  $1.55\pm0.03$ &  $6.37\pm0.03$ & $82\pm40$  &  3.0 & $924/850$ & 61\\
2 &  $1.60\pm0.03$ &  $6.41\pm0.04$ & $60\pm30$  & 3.8 & $992/991$ & 42\\
4 &  $1.70\pm0.02$ &  $6.41\pm0.02$ & $65\pm20$  & 3.5 & $1151/1194$ & 130\\
5 &  $1.72\pm0.03$ &  $6.41\pm0.03$ & $62\pm28$  & 3.6 & $990/1012$  & 67\\
6 &  $1.69\pm0.02$ &  $6.43\pm0.03$ & $58\pm20$  & 4.2 & $1175/1189$ & 101\\
\hline
\end{tabular}
\end{scriptsize}
\end{center}
\scriptsize (1) Emission line rest-frame energy centroid, in units of keV; (2) Emission line rest-frame equivalent width, in units of eV. 
The width of the line was fixed to $\sigma$=0.1 keV (see text for details); (3) 2--10 keV flux in units of 
10$^{-11}$ erg~s$^{-1}$~cm$^{-2}$\\
\end{table}

In a few observations (namely observations 4, 5 and 6) we found additional weak, but significant, 
residuals blueward of the 6.4 keV due to presumably Fe~K$\beta$ emission plus higher ionization iron 
emission components. Indeed, using this same dataset, together with a $\sim$100 ks Suzaku observation that 
was partially (for $\sim$25 ks) overlapping the observation n. 6, Ponti et al. (2009) have shown in a companion paper 
that this emission could be well reproduced by a broad line from an accretion disc, while it cannot be 
easily described by scattering or emission from photo-ionized gas at rest.
We did not investigate nor consider further these emission structures here, since the primary aim of the present
work is to perform a detailed search for absorption lines at energies above $\sim$7 keV. 
We checked nevertheless that different parameterizations and/or modelling of the complex 
Fe emision line structure, including the above weak ionized components, do not affect 
the results and conclusions presented here.\\

\subsection{Fe Absorption Lines}

Dadina et al.~(2005) reported the detection of both red- and blue-shifted
absorption lines due to highly ionized Fe in the \emph{BeppoSAX} 
and \emph{XMM-Newton} (non-simultaneous) spectra of Mrk~509. These lines were found at rest-frame energies 
of $\sim$5.5~keV and $\sim$8.2~keV, respectively. The first was seen in 
two \emph{BeppoSAX} observations, the latter was seen by both
satellites. Motivated by these results, we performed a systematic and blind search of 
absorption structures in the $\sim$4--10 keV band for the whole \emph{XMM-Newton} dataset available to date. 

We first fitted the 3.5--10.5 keV data with a simple power-law model
and searched for deviations in the $\Delta \chi^2$, as shown in the first and second panels of Figure 1 
for observation 1.
We then implemented a more systematic method to search for features in the spectrum, similar to the one 
introduced by Miniutti \& Fabian (2006) and Miniutti et al. (2007). 
We fitted the 3.5--10.5 keV data with the baseline power-law only (panel 3) or 
with the power-law plus Gaussian emission line model (panel 4) obtained 
in \S 2.2, stored the resulting $\chi^2$ value, and then freezed the baseline model parameters to their 
best-fit values. 
We then added a narrow (unresolved, $\sigma=0.01$ keV) Gaussian
line to the model, and then searched for the presence of both emission and 
absorption features by making a series of fits stepping the new Gaussian line 
energy in either the 4--10~keV or 7--10 keV intervals and its normalization between 
$-4\times10^{-5}$ and $+4\times10^{-5}$~ph~s$^{-1}$~cm$^{-2}$, each time
storing the new $\chi^2_i$ value. We then derived a grid of $\chi^2$
values and made a plot of the contours with same $\Delta\chi^2$ level with
respect to the baseline model fit.   
Negative $\Delta\chi^2$ values calculated in this way indicate that a better fit is reached with the 
inclusion of the line and -- like in the standard {\sf steppar}
type of contours in XSPEC,-- the $\Delta\chi^2$ values can be translated in a statistical
confidence level for the addition of 2 more parameters, as these do not depend on 
the sign but only on the absolute value of $\Delta\chi^2$. We then 
plot the standard values of $\Delta\chi^2$=-2.3, -4.61 and -9.21 relative to confidence levels of 68\%, 90\% and 
99\%, respectively. The major difference between these contours and the standard ones obtained 
with the {\sf steppar} command in XSPEC, is that the contours are here ``inverted'' in the sense 
that inner contours include higher significance than outer ones.
These contours are shown in panel 3 and 4 of Figure 1, for observation 1, 
as well as in Figures 2 and 3, for all the observations.

\begin{figure}[!]
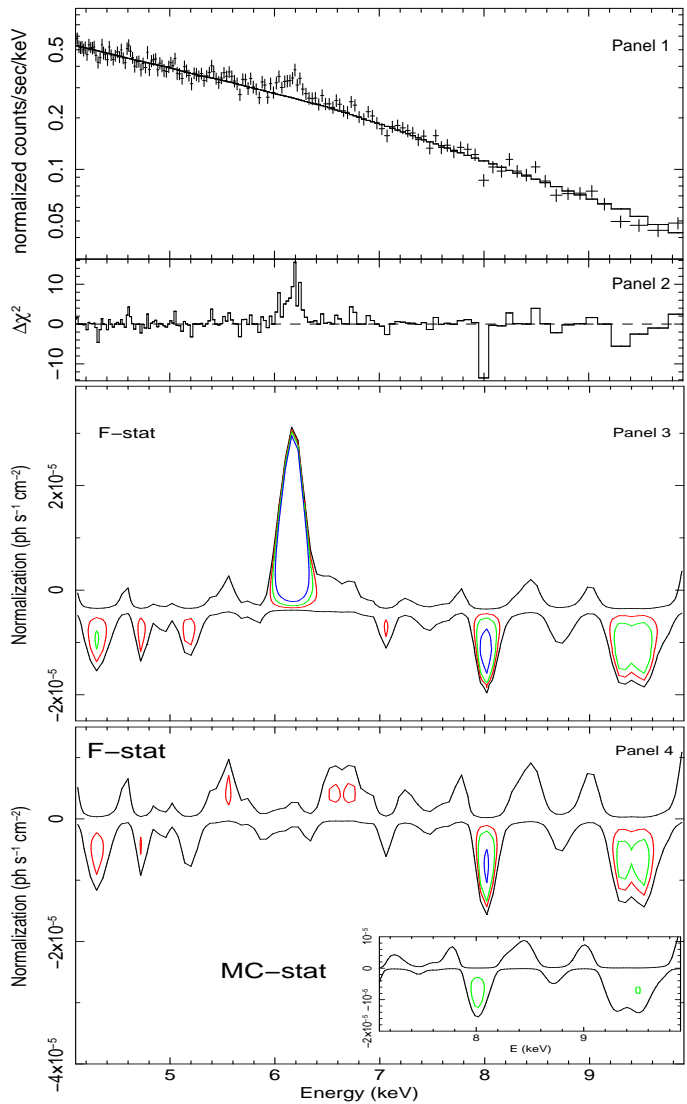

\parbox{7cm}{
\psfig{file=./plot_lda_chi_obs1.ps,width=9cm,height=5cm,angle=-90}}
\parbox{7cm}{
\vspace{-0.0cm}
\psfig{file=./plot_cont_4_10_obs1_fig1.ps,width=9cm,height=4.5cm,angle=-90}}
\parbox{7cm}{
\vspace{-0.00cm}
\psfig{file=./plot_merge_cont_4_10_fstat_mcstat_obs1.ps,width=9cm,height=5cm,angle=-90}}
\hspace{0.1cm} \
\caption{\emph{Panel 1}: EPIC pn spectrum of Mrk~509 (obs. 1) fitted in the 3.5--10.5 keV band, with data rebinned with S/N$>$12, fitted with a single power-law model; \emph{Panel 2}: $\Delta\chi^2$ residuals;  \emph{Panel 3}: F-statistics confidence contours (68\% (red), 90\% (green) and 99\% (blue) levels) with a single power-law model with Galactic absorption; \emph{Panel 4}: F-statistics confidence contours (68\%, 90\% and 99\% levels) after the inclusion of a Gaussian emission line in the model; and \emph{Panel 4, inset}: Confidence contours between 7--10 keV calculated using a Monte Carlo simulation (see text for details).
Contours (in black) calculated with $\Delta\chi^2$=0.5 are reported to indicate the continuum level.
}
\end{figure}

\begin{figure}[!]
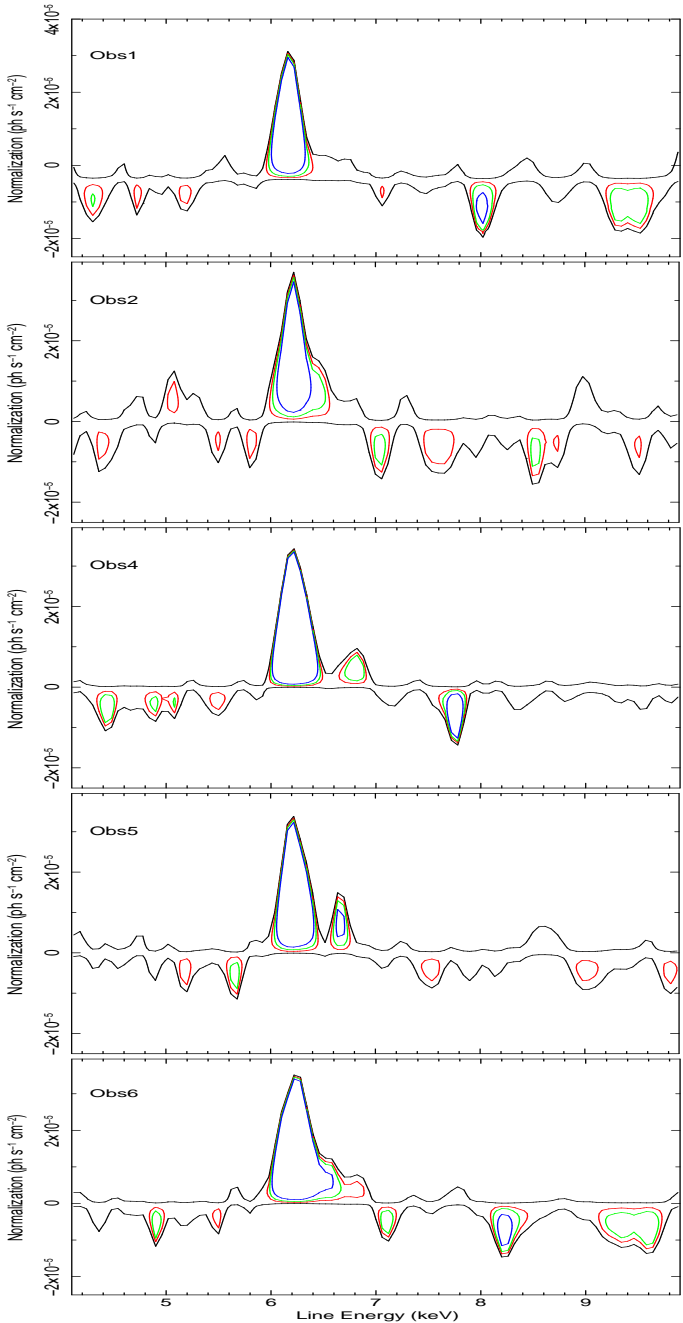

\parbox{7cm}{
\psfig{file=./plot_cont_4_10_obs1.ps,width=9cm,height=3.5cm,angle=-90}}
\parbox{7cm}{
\vspace{-0.0cm}
\psfig{file=./plot_cont_4_10_obs2.ps,width=9cm,height=3.5cm,angle=-90}}
\parbox{7cm}{
\vspace{-0.0cm}
\psfig{file=./plot_cont_4_10_obs3.ps,width=9cm,height=3.5cm,angle=-90}}
\parbox{7cm}{
\vspace{-0.0cm}
\psfig{file=./plot_cont_4_10_obs4.ps,width=9cm,height=3.5cm,angle=-90}}
\parbox{7cm}{
\vspace{-0.0cm}
\psfig{file=./plot_cont_4_10_obs5.ps,width=9cm,height=3.5cm,angle=-90}}
\caption{4-10 keV confidence contours for $\Delta\chi^2$=-2.3, -4.61 and -9.21 (i.e., 68\%, 90\% and 
99\% levels) calculated for both positive and negative residuals for a narrow Gaussian line 
in the data over the 4--10 keV energy range(see text for details). The contours at $\Delta \chi^2 = +0.5$ are also reported.}
\end{figure}

The probability levels shown in Figure 1 are derived from F-test only
statistics, and do not take into account the number of trials used by our line blind
search technique approach (e.g. Protassov et al. 2002). This might result
in an overestimation of the statistical significance associated with each line
feature. For this reason, contours derived with a method based 
on Monte Carlo simulations are shown for observation 1, in the inset of panel 4 of Figure 1.
We address issues and results from this method further in \S 2.4 below.

As for the redshifted absorption structures, we noted the presence in the data  
of weak deviations between 4 and 6 keV (see the residuals redward of the FeK emission line in Figure 2).
After inclusion of the neutral FeK emission line in the fit, we found that none of these absorption features 
is, however, statistically significant with upper-limits for an absorption line in that energy range 
of EW$\gsimeq$-15 eV (90\% confidence).
We did not confirm therefore the two positive BeppoSAX detections of the redshifted absorption line but
are still consistent with a scenario where the absorber is transient and/or patchy in nature, as proposed in 
Dadina et al. (2005).

\begin{table*}[!htb]
\caption{Mrk~509 absorption line parameters (rest-frame energies).}
\label{tab: mrk509_abs_par}
\begin{scriptsize}
\centering
\begin{tabular}{ c c c c c c c c}
\hline
N & E$^a$           & EW$^a$    & $\Delta\chi^2$ & F-test & MC$^b$ & Id & v/c\\
obs. & (keV)             &  (eV)          &                             &            &                        & (FeXXVI)& \\
\hline\\
1 & $8.29\pm0.05$  &  $-32\pm20$ & 9.6 & $\sim$99\% & 95\% &  K$\alpha$&0.172$\pm$0.005\\
    & $9.74\pm0.09$  & $-59^{+18}_{-22}$ & 8 & $\sim$98.5\% & 87.5\% & K$\beta$ & 0.162$\pm$0.009 \\
&\\
2 & $\equiv$8.29$^c$ &  $\gsimeq-23$ & $\dots$ & $\dots$ & $\dots$ \\
   & $\equiv$9.74$^c$ &  $\gsimeq-32$  & $\dots$ & $\dots$ & $\dots$ \\
&\\
4 & $8.03\pm0.03$&   $-20\pm10$ & 16.5 & $\gsimeq$99.9\% & 99.8\% &K$\alpha$&0.141$\pm$0.003\\
    & $\equiv$ 9.74$^c$  &  $\gsimeq-17$  & \dots & \dots & \dots\\
&\\
5 & $\equiv$8.29$^c$ &  $\gsimeq-14$ & $\dots$ & $\dots$ & $\dots$ \\
   & $\equiv$9.74$^c$ &  $\gsimeq-17$  & $\dots$ & $\dots$ & $\dots$ \\
&\\
6 & $8.50\pm0.05$  &  $-20\pm12$ & 10.2 & $\sim$99.5\% & 95.5\% &K$\alpha$&0.196$\pm$0.005\\ 
    & $9.8^{+0.1}_{-0.2}$  &  $-34^{+26}_{-29}$ & 7 & $\sim$98\% & 78\% & K$\beta$ &$0.17^{+0.02}_{-0.01}$\\
\hline\\ 
\end{tabular}
\end{scriptsize}
\\
\scriptsize{(a) Rest-frame energy. Errors are quoted at the 90\% and 68\% confidence for the first and second lines, 
respectively.  \\
(b) Significance calculated through Monte Carlo simulations. (c) Values frozen at the best-fit values obtained during obs. 1.}
\end{table*}

As for the blueshifted absorption structures, we find instead several in the data. 
Not only the previous \emph{XMM-Newton} detection during observation 1 by Dadina et al. (2005)
is confirmed, but the line is detected at a high significance level (see Table 3, and Figures 2 and 3) 
also during observations 4 and 6 (at E$\sim$8.2, 8.0 and 8.4 keV, respectively).
In addition, there is evidence during both observations 
1 and 6 that an additional absorption line, possibly broader or maybe due to a blend of lines
is present at even higher energies, at E$\sim$9.4 keV (observed frame), i.e. $\sim$9.7 keV 
(rest-frame). Confidence contours for these absorption lines are shown in Figures 2 and 3.  
We report in Table 3 the best-fit parameters for these lines, as well as the significance of their detections, 
calculated with the F-statistics (column 5) and the Monte Carlo simulations (column 6).
 
Remarkably, the two sets of two lines are consistent with the theoretical energy peaks of FeXXVI K$\alpha$ (6.966 keV) 
and K$\beta$ (8.268 keV), if blueshifted with respect to the systemic velocity by v$\sim$0.17c (obs. 1) 
and v$\sim$0.2c (obs. 6), respectively.
The line at $\sim$ 8 keV during obs. 4 is consistent with Fe XXVI K$\alpha$ at v$\sim$0.14c. 
If the lines were associated, instead, with FeXXV K$\alpha$ (6.697 keV) and K$\beta$ (7.880 keV), the requested 
velocity shifts would be consequently larger (i.e. v=0.21, 0.18 and 0.23~c for obs. 1, 4 and 6, respectively). 
 
We also checked the MOS results for consistency with the pn instrument. Remarkably, despite the considerably 
lower statistics for the MOS (due to the much lower effective area at E$>$7 keV), we found evidence, 
albeit with lower significance, of absorption features in all three observations 1, 4 and 6 and with energies and 
intensities consistent with the pn. 
In particular, we found statistical improvements by $\Delta \chi^2$=13, 5 and 3, for the addition of a line at 
E$\sim$8.2, 8.0 and 8.4 keV, with EW$\simeq$-67$\pm$30 eV, -24$\pm$23 eV, and -17$\pm$16 eV (at 90\% confidence for one additional d.o.f.), 
respectively.
These results suggest these lines are not statistical fakes (but see also \S 2.4).

The pn background spectrum is known to have two instrumental emission features, one at observed energy of 
$\sim$7.5 keV due to a Ni K$\alpha$ emission line and one at $\sim$8 keV due to a Cu K$\alpha$ emission line 
(Katayama et al. 2004, Freyberg et al. 2004). These may have a strong impact on the source high-energy spectrum.
We checked however that this is not the case for Mrk~509. The source is so bright (F(2-10 keV)$>$2$\times$10$^{-11}$erg cm$^{-2}$s$^{-1}$) 
that the background level was always negligibly small, even in the FeK bandpass and for both pn and MOS (see Figure 4), 
thereby excluding a possible artifact due to the instrumental background. 

During observations 2, 4 (for the $\sim$9.7 keV line) and 5 where the features were not detected, we obtained upper-limits 
on the absorption lines at energies of E$\simeq$8.3 and 9.7 keV, with EW$\gsimeq$-15 and $\gsimeq$-30 eV, respectively (see Table 3).


\begin{figure}[!]
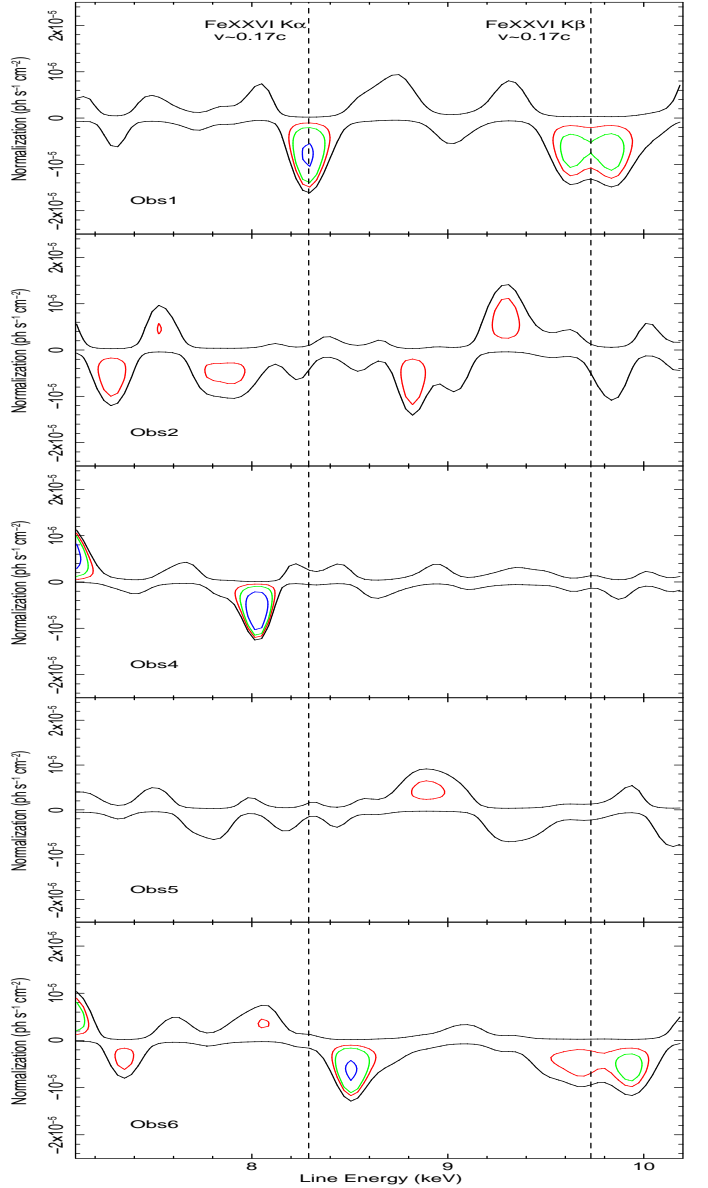

\parbox{7cm}{
\psfig{file=./plot_cont_7_105_lines_obs1.ps,width=9cm,height=3.1cm,angle=-90}}
\parbox{7cm}{
\vspace{-0.05cm}
\psfig{file=./plot_cont_7_105_lines_obs2.ps,width=9cm,height=3.1cm,angle=-90}}
\parbox{7cm}{
\vspace{-0.05cm}
\psfig{file=./plot_cont_7_105_lines_obs3.ps,width=9cm,height=3.1cm,angle=-90}}
\parbox{7cm}{
\vspace{-0.05cm}
\psfig{file=./plot_cont_7_105_lines_obs4.ps,width=9cm,height=3cm,angle=-90}}
\parbox{7cm}{
\vspace{-0.05cm}
\psfig{file=./plot_cont_7_105_lines_obs5.ps,width=9cm,height=3.5cm,angle=-90}}
\hspace{0.1cm} \
\caption{Zoomed, rest-frame 7-10 keV confidence contours for $\Delta\chi^2$=-2.3, -4.61 and -9.21 (i.e., 68\%, 90\% and 
99\% levels, colors are as those in Fig. 1) calculated for both positive and negative residuals for a narrow Gaussian line 
in the data between 7--10 keV (see text for details).
The contours at $\Delta \chi^2 = +0.5$ are also reported.}
\end{figure}

\begin{figure}[!]
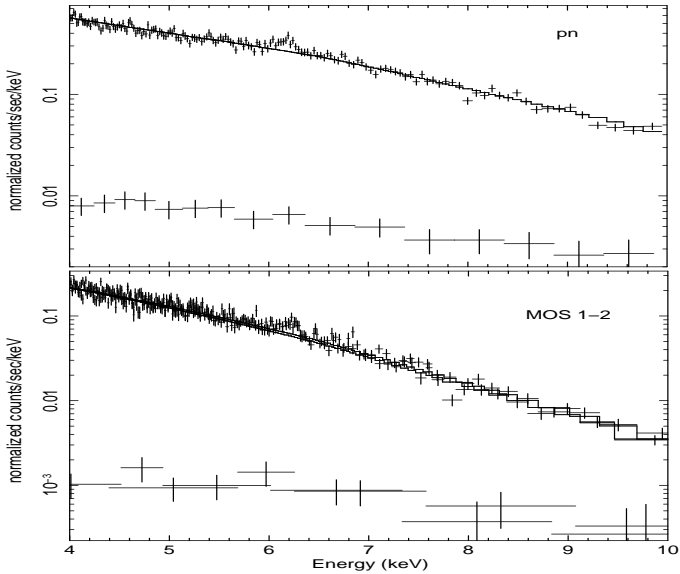

\parbox{7cm}{
\psfig{file=./plot_pn_back_obs1.ps,width=8.8cm,height=3.5cm,angle=-90}}
\parbox{7cm}{
\psfig{file=./plot_mos12_back_obs1.ps,width=9.1cm,height=4cm,angle=-90}}
\caption{pn source and background during obs. 1 (Top), rebinned with $S/N>12$; MOS source and background 
during obs. 1 (Bottom), rebinned with $S/N>6$ and 3, respectively. The spectra show that the source 
is over an order of magnitude higher in counts s$^{-1}$ keV$^{-1}$ compared to the backgrounds 
in the FeK bandpass, in both pn and MOS.
}
\end{figure}

Finally, there are indications that the lines do vary among the observations (see Figure 3, where contours are plotted at 
rest-frame energies), clearly in blueshift/energy 
(being 0.17c, 0.14c and 0.20c during obs. 1, 4 and 6 respectively) and possibly in intensity, since they were not 
detected during observations 2 and 5. In particular, albeit marginal, the intensity variations are found on 
a time-scale as short as $\sim$ 100 ks, i.e. the time encompassed between obs. 4 and 5, but one should recall 
the intrinsic difficulty to place here stringent upper-limits on the EW of lines for which the 
velocity is unknown a-priori. The line velocity variations are instead unambiguous (see Figure 3), 
on a time-scale of $\sim$ 6 months, i.e. the time encopassed between obs. 4 and obs. 6.
Given the mass of the black hole in Mrk~509 of M$_{bh}$ $\sim$ 1.4 $\pm$0.1 $\times$10$^{8}$ M$_{\odot}$
(Peterson et al. 2004), such variability time-scales suggest the presence of important 
radial or transverse motions quite close to the central supermassive black hole as discussed below. 

\subsection{On the statistical significance of the absorption lines}

The F-test detection probabilities derived here (see Table 3) from our line blind search 
can be overestimates of their real values (e.g. Protassov et al.,  2002). We therefore applied
a more rigorous method that uses extensive Monte Carlo simulations
(see e.g. Porquet et al.~2004; Miniutti \& Fabian 2006; Markowitz et al.~2006)
 in order to assess the detection probabilities from our line blind search.
For all three cases in which one or two lines have been detected (i.e. observations
1, 4 and 6) we considered the absorbed power law plus narrow Gaussian
emission line as the null-hypothesis model. 

We simulated a null-hypothesis model spectrum with the same exposure as
the real data and subtracted the appropriate background. 
This spectrum was fitted again with the null-hypothesis model in
the \mbox{3.5--10.5}~keV band and the new parameters were recorded. Then,
we generated a new simulated spectrum with this refined null-hypothesis
model in order to account for the uncertainties in the model itself.
Therefore, we made a fit of the null-hypothesis model to this new
simulated spectrum, considering only the 3.5--10.5~keV band and recorded
the $\chi^2$ value. We then added a narrow Gaussian line ($\sigma=0.01$~keV)
to the model, with its normalization free to be positive or negative  
and stepped the centroid energy between 7 and 10 keV at intervals of 100~eV. 
Each time we made a new fit and recorded the maximum of these 
simulated $\Delta\chi^2$.
Then, we repeated the above procedure $S=10^4$ times in order to 
generate a distribution of $10^4$ simulated $\Delta\chi^2$ values. If $N$  
simulated $\Delta\chi^2$ values were greater or equal to the real one, then the 
detection confidence level corresponds to $1-N/S$.\\

The new values of line detection confidence levels based on the above Monte Carlo 
simulations are reported in Table 3, in comparison to the F-test probabilities. 
We also estimated the combined Poissonian probability of detecting by chance the 3 absorption lines 
(at E$\sim$8-8.5 keV) in 5 observations, and this turned out to be $\leq$0.001
(i.e. a probability $\geq$99.9\% that these lines are not fake).
Considering also the two lines at E$\sim$9.7-9.8 keV, and that they turn out to be 
at the expected energies for K$\beta$ transitions (i.e. the F-statistics alone should properly work 
for these), would lower the 
combined probability even further. For example, a fit with the two lines (FeXXVI K$\alpha$ and K$\beta$) 
with their energies freezed at their theoretically expected values, and same blueshifted velocity, yielded 
a  $\Delta\chi^2$ improvement of 19 and 13, for 3 interesting parameters, corresponding 
to a probability of $>$99.9\% and 99.5\% during observations 1 and 6, respectively.

\begin{figure}[!]
\parbox{7cm}{
\psfig{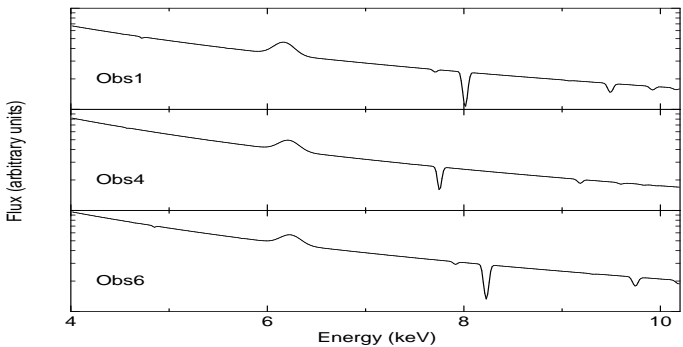}}
\caption{Best-fit photoionized models for observations n.1 (top), 4 (middle) and 6 (bottom). See text for details.}
\end{figure}

\section{Discussion and Conclusions}

The main results of our spectral analysis are the following:

\begin{itemize}

\item We detected several absorption features in the energy range $\sim$8-10 keV in three 
(out of six) \emph{XMM-Newton} observations of the bright and luminous Seyfert 1 galaxy Mrk~509. 
The equivalent widths of the lines are in the range between -20 and -60 eV.

\item The most likely identification for these features being 
FeXXVI K$\alpha$ and K$\beta$, they imply (mildly) relativistic 
blueshifts in the range $\sim$0.14-0.2c, and low K$\alpha$/K$\beta$ ratios, of 
the order of unity.

\item There is evidence for line velocity variations, as well 
as for appearance and disappearance of the lines, on time-scales as 
short as $\sim$100 ks. 

\end{itemize}

The above results are reminiscent of the Seyfert 1.5 NGC1365 (Risaliti et al. 2005), in particular 
for i) the identification of the absorption lines 
with FeXXVI K$\alpha$ and K$\beta$ lines, ii) the small K$\alpha$/K$\beta$ ratios, and iii) the 
velocity variations detected on similarly short time-scales for a comparable black hole mass.
Differences are that here we only detect one ionization state for iron (instead of two, FeXXV and FeXXVI, 
for NGC1365), and the line velocities are here about ten times larger than in NGC1365. As shown below, this could be due to 
an intrinsically more extreme outflow in Mrk~509 with respect to NGC1365, in terms of either velocity or ionization state, or both. 
Alternatively, the structure and velocity field of radiatively-driven winds being a function of the 
polar angle (e.g. Proga \& Kallman 2004), a different viewing angle as expected for their type 1 (for Mrk509) and 
type 2 (NGC1365) classification would naturally imply different average velocities along the line of sight.

Outflowing material is expected to imprint a characteristic P-Cygni profile in the data, where the detailed 
absorption and re-emission line profile depends strongly on the flow/wind geometry and density. Theoretical 
predictions based on Monte Carlo calculations have been presented in, e.g., Schurch \& Done (2007) and 
Sim et al. (2008), with some recent attempts to fit the best available observational data (Sim et al. 2008, Pounds 
\& Reeves 2008).
Rather than attempting here such a full treament and modelling of the absorber
with data of admittedly too low statistics for such a purpose, we discuss the above results 
in the light of simpler and more qualitative arguments.

Following Risaliti et al. (2005) and from the curves of growth presented in Bianchi et al. (2005), an equivalent 
width of the K$\alpha$ lines EW$\sim$20-30 eV can be obtained with N$_{\rm H} \gsimeq$5$\times$
10$^{23}$ cm$^{-2}$ and log$\xi$$\gsimeq$4 erg cm s$^{-1}$, with turbulent velocity set to zero. 
With larger values of turbulence,  i.e. velocities between 100 and 1000 km/s, the column density 
needs not to be as large, with values of N$_{\rm H} \sim$5-10$\times$10$^{22}$ cm$^{-2}$.
These values would also be consistent with the low K$\alpha$/K$\beta$ ratios, which indicate that saturation effects 
are significant (see Figure 4 in Risaliti et al. 2005).
Given the low statistics available, velocity widths as large as a few thousands km/s, i.e. consistent 
with the above estimates, can not be excluded here. Such high values of turbulence 
would also be reasonable for bulk motions in an outflow with such large velocities.

These rough estimates were then confirmed through a more detailed, but still approximate, fitting 
of the FeK absorption lines with two different phenomenological models: \emph{zxipcf}, a partial covering 
of photo-ionised absorbing material that is incorporated into the XSPEC package (Reeves et al 2008), 
and XSTAR tables calculated using an input SED with a power-law photon index of $\Gamma$=2
and a turbulent velocity of 1000 km/s. We obtained good and consistent fits to the data. 
For all three observations (obs. 1, 4 and 6), the best-fit parameter values were log$\xi$$\sim$5$\pm$0.4 erg cm s$^{-1}$, 
N$_{\rm H} \sim$2-4$\times$10$^{23}$ cm$^{-2}$ and a blueshift of v$\sim$0.14-0.2c. 
Best-fit models obtained with the XSTAR photoionization model are shown in figure 5.
In these models, the lines are produced by only FeXXVI (K$\alpha$ and K$\beta$) and, despite the large columns, 
no other signature is expected at lower energies because of the extremely large ionization.

Another possible solution, albeit slightly statistically worse, could also fit the EPIC data with 
less extreme values of N$_{\rm H} \sim$10$^{22-23}$ cm$^{-2}$ and log$\xi$$\sim$3 erg s$^{-1}$ cm. In this case, the 
absorption lines are identified with mainly FeXXV. This solution does, however, require even larger velocities 
up to 0.18-0.23c. Moreover this solution appears to be slightly fine-tuned because a slightly lower value of 
$\xi$ would imply a significant low-energy curvature and a higher value of $\xi$ would give rise to lines from 
also FeXXVI. We also checked but did not find any evidence in the RGS data of OVIII absorption lines at the 
above mentioned velocities, the data yielding quite stringent limits of EW$\lsimeq$0.2-0.4 eV. Altogether, this 
solution appears to be less likely than the one based on FeXXVI ``only'' and was thus not considered any further 
here.

The high ionization parameter, together with the variations seen in the 
absorption lines can be used to infer limits on the distance and size of
the absorber. Assuming that the ionization state is due to illumination from the central 
X-ray and UV source, rather than heated/illuminated locally, 
we can use $\xi$=$L_{ion}$/(nr$^{2}$), where n is the number density, $L_{ion}$
the source continuum luminosity (integrated between 1 and 1000 Rydberg) to estimate the maximum distance $r$
of the absorber from the illuminating source.
The observed line-of-sight absorbing column $N_{\rm H}$ is a function of the density n of the material at a ionisation 
parameter $\xi$ and the shell thickness $\Delta r$: $N_{\rm H} = n{\Delta}r $.
Making the reasonable assumption (given the r$^-\alpha$ dependencies of winds/outflows) that the thickness $\Delta r$ 
is less than its distance from the source $r$, and combining the expression for 
the ionisation parameter, we obtain the limit $r< L_{ion}/(\xi N_{\rm H})$. Using $L_{ion}$=2$\times$10$^{44}$ erg/s, 
$N_{\rm H}$=10$^{23}$~cm$^{-2}$ and log$\xi$$\sim$5 erg~cm~s$^{-1}$, we obtain r$\leq$ 2$\times$10$^{16}$~cm. 
Assuming the black hole mass of M$_{bh}$ = 1.4 $\times$10$^{8}$~M$_{\odot}$ (Peterson et al. 2004) for Mrk~509, this 
corresponds to a location for the absorber within a distance $<$500 Schwarzschild radii (r$_{s}$) from the black hole.

Interestingly, we note that the variability of the absorption lines is indicative of a compact absorber and rules 
out a local origin, from either the Galaxy or the Local Group, for the absorber (McKernan et al. 2004, 2005). 
In fact the absorber appears to be even more compact than the broad line region which has an estimated size of the order of 
$\sim$ 80 light-days (Kaspi et al. 2005). 
Analogous short term variability of iron absorption lines velocity and
intensity was recently found for the BAL QSO APM 08279+5255 (Saez et al.
2009), suggesting that rapid changes in the geometry of the outflow and/or
of physical parameters of the absorber are a fundamental characteristic of
the X-ray component of AGN winds.

A rough estimate of the escape velocity along the distance of a Keplerian disk is given by the equation 
$v^{2} = 2 GM/r$, which can be written as $v=(r_{s}/r)^{1/2}c$. The observed blueshift velocity of the absorber
v$\sim$0.14-0.2c is therefore much larger than the escape velocity at $\sim$ 500 r$_{s}$, of the order 
of $\sim$0.05c, as well as the escape velocity at 100 r$_{s}$, that is $\sim$0.1c.
Assuming a constant velocity for the outflow and the conservation of the total mass, we can thus estimate the mass loss rate $\dot M_{out}$ 
associated to this outflow as $\dot M_{out} \sim \Omega n r^{2}vm_{p}$, where $v$ is the outflow velocity, n is the ion number density 
of the absorber, $m_p$ is the proton mass and $\Omega$ is the covering fraction of the outflow. 
Following Blustin et al. 2005, this translates to $\dot M_{out} \sim$ 1.23 $m_{p} {L_{ion}\over{\xi}} \Omega v$, i.e. for 
Mrk509, to $\dot M_{out}$ $\lsimeq$ 2.4 $\times$ ($\Omega$/4$\pi$) $v_{0.1}$ M$_{\odot}$ yr$^{-1}$, with $v_{0.1}$ the flow 
velocity in units of 0.1 c. This corresponds then to a kinetic power 
of ${1\over{2}}$$\dot M_{out} v^2$ $\lsimeq$ 6.7$\times$ 10$^{44}$ ($\Omega$/4$\pi$) $v_{0.1}$$^3$ erg s$^{-1}$ that could be injected 
into the ISM.

Following McKernan et al. (2007), we can derive a simple relation for the ratio between the mass outflow rate and the rate 
of the accretion flow, i.e. $\dot M_{out}$/$\dot M_{acc}$ $\sim$ 6000 ($\Omega$/4$\pi$)($v_{0.1}$/$\xi_{100}$)$\eta_{0.1}$, where  
$\Omega$ is the absorber covering fraction, $\xi_{100}$ is the ionization parameter in units of 100 erg cm s$^{-1}$ and 
$\eta$=$\eta_{0.1}$$\times$0.1 is the accretion efficiency of the black hole.
For Mrk~509, this relation yields $\dot M_{out}$/$\dot M_{acc}$ $\sim$10 ($\Omega$/4$\pi$), implying either a low covering 
fraction and/or an intermittent outflow to permit the growth of the black hole.
Viceversa, even a moderately large covering factor for the ouflow would imply significant and energetically relevant 
mass losses into the ISM.

The high velocity, high ionization and variability properties of this absorber 
suggest that it may be associated with an outflow launched by an accretion disk 
at relatively small radii of less than a few hundred Scwarzshild radii.
This adds to the long and well known evidence from the soft X-rays of a multi-phase warm absorber in Mrk~509 
with log$\xi$$\sim$0.9-3.26 and $N_{\rm H}$$\sim$0.8-5.8$\times$10$^{21}$ cm$^{-2}$ 
(Smith et al. 2007,), and to the ``intermediate'' (log$\xi$$\sim$5.4, $N_{\rm H}$$\sim$5.8$\times$10$^{22}$ cm$^{-2}$ and v$\sim$0.05c) 
absorption component found by Ponti et al. (2009) based on continuum variability arguments. 
Overall, Mrk~509 shows compelling evidence for a stratified 
absorber, as those predicted by hydrodynamical simulations of radiatively and/or magnetically 
driven outflows (Proga 2003, Proga \& Kallman 2004, King \& Pounds 2003, Murray et al. 1995), but with ionizations 
and velocities up to extreme and challenging values. 
The extreme ionization parameter raises the question as to how much more fully ionised matter is included, but undetected yet, 
in the outflow, with its implications for the outflow mass-loss rates and energetics.

Finally, the frequency in AGNs of such extreme outflows is to date unclear.
Proper statistical studies on larger, complete, samples are being performed (Tombesi et al., in preparation). They are needed 
to secure the statistical significance of these features, and to assess their frequency and typical characterisation in terms of 
density, velocities, ionization states and covering factor. Unfortunately current available data do suffer an 
``observational bias'' against the detection of the highest-velocity blueshifted features in that 
orbiting X-ray telescopes are of limited spectral and sensitivity capabilities at  E$>$7 keV. 
Future missions such as $IXO$ will likely allow an exciting step forward in studying mildly relativistic 
outflows in AGNs, as demonstrated by the first attempted simulations (Cappi et al. 2008, Tombesi et al. 2008). 

\begin{acknowledgements}

This paper is based on observations obtained with the \emph{XMM-Newton} 
satellite, an ESA funded mission with contributions by ESA Member States and 
USA. We would like to thank the referee K. Pounds for his constructive suggestions.
We thank A. De Rosa for usefull discussions. MC, MD and GP acknowledge financial support 
from ASI under contracts ASI/INAF I/023/05/0 and I/088/06/0. GP aknowledges ANR for support 
under grant number ANR-06-JCJC-0047.

\end{acknowledgements}

{}

\end{document}